\documentclass[conference]{IEEEtran}
\IEEEoverridecommandlockouts

\usepackage[utf8]{inputenc}
\usepackage{cite}
\usepackage{amsmath,amssymb,amsfonts}
\usepackage{algorithmic}
\usepackage{array}
\usepackage{graphicx}
\usepackage{textcomp}
\usepackage{xcolor}
\usepackage{multirow}
\usepackage{makecell}
\usepackage{hhline}
\usepackage{dirtytalk}
\usepackage{eso-pic}
\usepackage{url}
\usepackage[hidelinks]{hyperref}

\newcolumntype{L}{>{\centering\arraybackslash}m{3cm}}

\def\BibTeX{{\rm B\kern-.05em{\sc i\kern-.025em b}\kern-.08em
		T\kern-.1667em\lower.7ex\hbox{E}\kern-.125emX}}

\definecolor{orcidlogocol}{HTML}{A6CE39}
\newcommand{\orcidicon}[1]{%
	\href{https://orcid.org/#1}{%
		\textsuperscript{%
			\raisebox{-0.2ex}{\includegraphics[height=1.4ex]{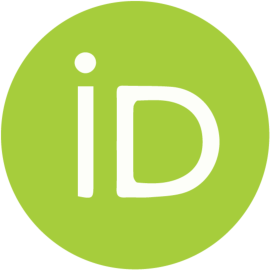}}%
		}%
	}%
}

\title{Controlled Out-of-Band Device-to-Device Communication in Cellular Networks Using a Backup Channel in Television White Space}

\begin{document}
	
	\author{
		\IEEEauthorblockN{
			Saifur Rahman\IEEEauthorrefmark{1}\,\orcidicon{0000-0002-7262-183X},
			Syed Luqman Shah\IEEEauthorrefmark{2}\,\orcidicon{0000-0001-5792-0842},
			Salim Nasar Faraj Mursal\IEEEauthorrefmark{1},
			Ziaul Haq Abbas\IEEEauthorrefmark{2}\IEEEauthorrefmark{4}\,\orcidicon{0000-0003-4466-0089},\\
			Muhammad Usman\IEEEauthorrefmark{5},
			Muhammad Irfan\IEEEauthorrefmark{1}, and
			Fazal Muhammad\IEEEauthorrefmark{2}\IEEEauthorrefmark{3}\,\orcidicon{0000-0003-0405-0083}
		}
		\IEEEauthorblockA{\IEEEauthorrefmark{1}Electrical Engineering Department, College of Engineering, Najran University, Najran 61441, Saudi Arabia}
		\IEEEauthorblockA{\IEEEauthorrefmark{2}TeleCoN Research Lab, GIK Institute of Engineering Sciences and Technology, Topi 23640, Pakistan}
		\IEEEauthorblockA{\IEEEauthorrefmark{4}Faculty of Electrical Engineering, GIK Institute of Engineering Sciences and Technology, Topi 23640, Pakistan}
		\IEEEauthorblockA{\IEEEauthorrefmark{3}Department of Electrical Engineering, University of Engineering and Technology, Mardan 23200, Pakistan}
		\IEEEauthorblockA{\IEEEauthorrefmark{5}Department of Computer Science, University of Engineering and Technology, Mardan 23200, Pakistan}
		\IEEEauthorblockA{\IEEEauthorrefmark{3}Corresponding author: fazal.muhammad@uetmardan.edu.pk}
	}
	
	\maketitle
	
	\begin{abstract}
		In this article, we address the problem of spectrum scarcity in cellular networks (CNs). We propose a backup channel (BuC) for cellular users (CUs) located in the same macro-cell under the control of a single macro base station (eNB). This BuC operates in television white space and is detected by the CUs through a cognitive radio energy-detection channel-sensing technique with a certain probability of success. When all regular channels with the cellular eNB are occupied, the CUs within the same coverage area of the macro eNB can utilize the sensed BuC to establish a controlled out-of-band device-to-device link for communication. The BuC bypasses the eNB for data communication and reduces the burden on the core of the CN. This leads to improved cellular eNB capacity. In the proposed system model, each CU and eNB is equipped with two antennas for communication in two separate bands, i.e., cellular and TV bands. Simulations show significant reductions in the blocking probability and probability of call delay.
	\end{abstract}
	
	
	\section{Introduction}
	The basic assumption behind cellular networks (CNs) is to boost network capacity and exploit the power falloff with propagation distance, known as attenuation. Specifically, in CNs, a given area is split into smaller areas called clusters, each with its own set of channels $C_c$. A cluster is a group of cells, each cell has a macro base station (eNB), and $C_c$ is divided among the eNBs of the cells in the cluster. In another cluster at a different location, the same channel set $C_c$ is reused. This concept is called channel reuse or frequency reuse~\cite{mishra2007advanced}.
	
	Frequency channel sets $C$, where $C$ is a subset of $C_c$, are assigned to the eNB of a cell. When a cellular user (CU) wants to initiate a call, it requests service from the eNB by providing the necessary information, such as its location, mobile identification number (MIN), international mobile equipment identity (IMEI) number, subscriber identity module (SIM) number, etc. The eNB forwards this information to the mobile switching center (MSC) because the network provider must locate the other user (callee) and authenticate both CUs, i.e., the caller and the callee. The MSC locates the callee and asks the eNB to assign free channels to the CUs for their communication. If the eNB has no vacant channel available, the CU request is blocked or queued (delayed). Moreover, a smooth handoff is needed when a mobile CU moves from one cell region to another. The ongoing call is dynamically transferred from one cell's eNB to another cell's eNB during the handoff~\cite{anwar2023handoff}. The handoff call is blocked or delayed if there is no free channel available with the new eNB.
	
	Blocking or delaying handoffs and new calls degrades the performance of CNs in congested and urban environments. To mitigate these issues, two potential solutions have been proposed in the literature: (i) heterogeneous cellular networks (HetCNets)~\cite{stanic2023survey} and (ii) device-to-device (D2D) communication~\cite{gismalla2022survey}. HetCNets enable flexible deployment using a mix of macro, pico, and femto cells, each with its own relay access point (AP). These low-power APs eliminate coverage holes in the macro-cell and improve capacity in hotspots. However, deploying several APs in each micro-cell is costly, and managing interference becomes more complex. To reduce interference and cost relative to HetCNets, D2D communication has been proposed for rural and suburban areas. D2D communication enables two nearby users to communicate directly without the involvement of the eNB. Thus, it can reduce transmission delay, offload traffic from the eNB, increase network spectral efficiency and energy efficiency, and alleviate congestion in cellular core networks.
	
	However, in densely populated urban areas, marketplaces, international sports stadiums, and traditional or regional events, call delay and blocking become serious issues. These events are mostly limited to the range of a single macro-cell and the coverage of a single eNB. To improve the quality of service (QoS) in such scenarios, we propose establishing a D2D link in television white space (TVWS).
	
	TVWS refers to TV channels in the ultra-high-frequency (UHF) range that are not in regular use for TV broadcasting. In the United States (US), this frequency range extends from 54 MHz to 698 MHz. On September 23, 2010, the US Federal Communications Commission (FCC) issued a landmark order~\cite{FCCWhiteSpaces}, which allowed unlicensed users to use UHF white spaces that were formerly reserved for TV. An updated version of this FCC order was released on October 28, 2020~\cite{FCCWhiteSpacesDoc}.
	
	This work utilizes the useful characteristics of TVWS, such as spectrum availability, wider bandwidth, favorable propagation characteristics, longer propagation range, and spectrum fragmentation~\cite{arteaga2022toward}. The authors in~\cite{usman2015energy} proposed a BuC in cognitive radio networks (CRNs), in which a parallel channel is sensed along with the operating channel by secondary users (SUs). This work extends the BuC concept to CNs.
	
	\subsection{Novelty and Contribution}
	We introduce a BuC that is sensed by CUs in the TV spectrum. Whenever a call request is generated from a CU, instead of blocking or delaying the call due to the unavailability of a vacant channel in the cellular band, the eNB checks whether both CUs (i.e., caller and callee) are in the same macro-cell. If both the caller and callee are in the same cell, they can connect directly through the sensed BuC under the control of the eNB. The proposed model is beneficial in congested environments such as remote campuses, sports stadiums, marketplaces, and disaster-relief scenarios. The novelty and major contributions of this paper are as follows:
	
	\begin{itemize}
		\item A BuC for CUs in TVWS is proposed to improve eNB capacity and QoS.
		\item The proposed scheme shifts the communication paradigm from higher frequencies (cellular band) to lower frequencies (TV band), since high-frequency signals are heavily affected by noise and path loss.
		\item The proposed scheme provides network service for CUs that want to communicate with other CUs in the same cell in congested and urban environments, without delaying the call or placing it in a queue.
	\end{itemize}
	
	The remainder of this article is structured as follows. In Sec.~\ref{Sec3: System Model}, the system model is explained, followed by the proposed work and problem solution in Sec.~\ref{Sec4: Proposed Work}. In Sec.~\ref{Sec5: Simulation Results}, we present the simulation results and discuss the obtained results. Finally, in Sec.~\ref{Sec6: Conclusion}, we conclude the paper and suggest relevant future research directions.
	
	\section{System Model}\label{Sec3: System Model}
	This section presents the proposed network model and provides a concise overview of D2D communication and the CR energy-detection channel-sensing technique.
	
	\begin{figure}[t]
		\includegraphics[width=\linewidth]{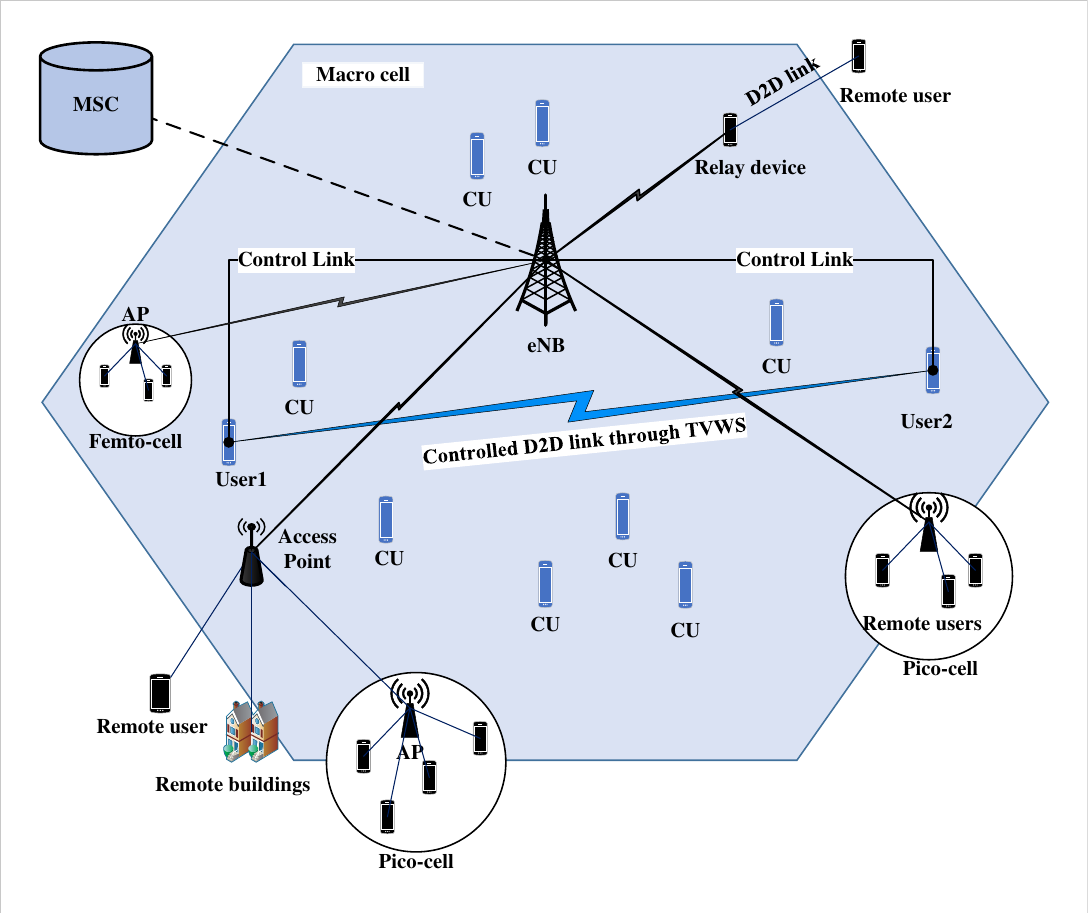}
		\caption{Proposed system model.}
		\label{Fig1: System Model}
	\end{figure}
	
	\subsection{Network Model}
	In our system model, a single macro-cell from a cluster of CNs is considered. A user named \say{User1} initiates a call to another user, \say{User2}. To initiate the call, User1 sends all necessary information, such as the MIN, location, IMEI, and SIM number of User2, to the eNB and requests a vacant channel. The eNB forwards this information to the MSC for user authentication and to find the location of User2. Simultaneously, during this process, User1 performs spectrum sensing in the TVWS using a CR channel-sensing technique. The success of this sensing process is indicated by the probability of success $P_d$. If User1 successfully detects a free channel in TVWS, it informs its eNB about the detected BuC. If User1 and User2 are situated within the same macro-cell under the control of the same eNB, and if there are no vacant channels available with their eNB, they can still communicate using the sensed BuC under full eNB control, as shown in Fig.~\ref{Fig1: System Model}. This communication mode is called \say{device-to-device (D2D) communication with operator control (DC-OC)}. For simplicity, we employ energy detection as the CR channel-sensing method in our research model. However, the eNB retains control over crucial aspects such as data transmission rate, power management, resource allocation, security, interference management, and other necessary processes. The eNB serves as the control link, ensuring the efficient and effective operation of these processes.
	
	\subsection{D2D Communication}
	In D2D communication, two nearby devices can communicate with each other without the involvement of the eNB, which reduces the load on the eNB and enhances cell capacity in CNs. The following potentials and advantages make D2D communication one of the key enabling technologies for 5G and beyond-5G networks~\cite{gismalla2022survey}: spectrum reuse, potentially improved throughput through high data rates, lower error rates, energy efficiency or optimal power consumption, low latency, fairness, and coverage over the entire cell. However, D2D communication users encounter several challenges, such as interference management, power control, and device discovery.
	
	The classification of D2D communication is given in Table~\ref{Table:1}. This work considers communication between any two nearby CUs as controlled D2D communication.
	
	\begin{table*}[htbp]
		\caption{D2D communication classification based on eNB involvement}
		\centering
		\renewcommand\arraystretch{2}
		\begin{tabular}{|>{\centering\arraybackslash}m{2.75cm}|>{\centering\arraybackslash}m{7cm}|>{\centering\arraybackslash}m{6cm}|}
			\hline
			\textbf{Technology} & \textbf{Categories} & \textbf{Subcategories}\\
			\hline
			\multirow{4}{2.75cm}{\centering\textit{D2D communication}: Data exchange between two devices without the aid of an eNB.}
			& \multirow{2}{7cm}{\centering\textit{Inband D2D communication}: D2D users use shared cellular networks and licensed frequency spectrum for communication.}
			& \textit{Underlay}: Cellular users and D2D users share some spectral resources.\\ \hhline{~~-}
			& & \textit{Overlay}: In this type of D2D communication, CNs have more channel resources; thus, interference between CUs and D2D users is prevented by allocating dedicated resources orthogonally to D2D users and CUs.\\ \hhline{~--}
			& \multirow{2}{7cm}{\centering\textit{Out-of-band D2D communication}: D2D users use unlicensed frequency spectrum for communication. The communication can be fully controlled by the eNB, or it can be managed by the device itself.}
			& \textit{Controlled D2D}: The communication is fully controlled by the eNB.\\ \hhline{~~-}
			& & \textit{Autonomous}: The D2D users manage the communication themselves.\\
			\hline
		\end{tabular}
		\label{Table:1}
	\end{table*}
	
	\subsection{Energy-Detection Sensing Technique}
\begin{figure}[htbp]
		\includegraphics[width=\linewidth]{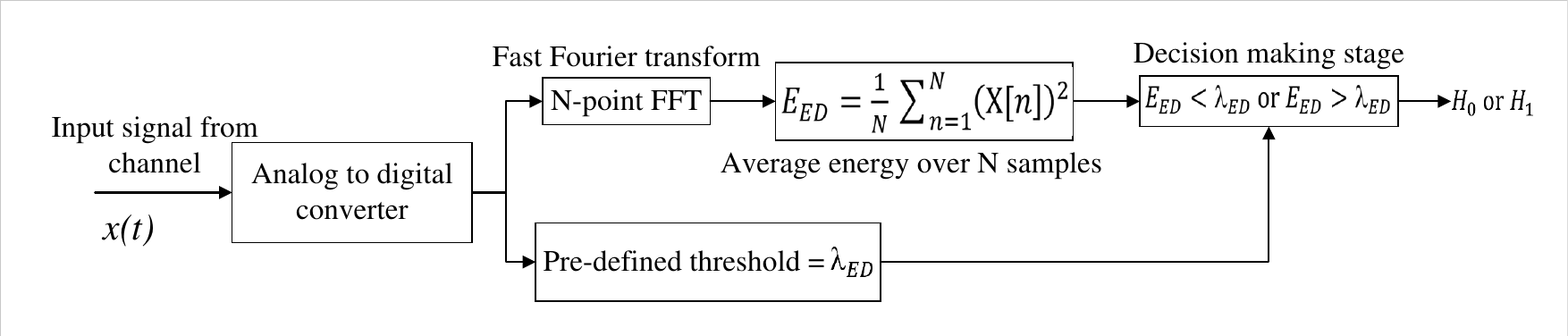}
		\caption{Energy-detection channel-sensing technique.}
		\label{Fig2: Energy detection}
	\end{figure}

	CR is a software-defined radio that automatically and intelligently detects the surrounding radio frequencies and adapts its operating parameters to the network infrastructure while meeting user demands~\cite{arshid2022energy, luo2022energy}. Energy detection is one of the CR narrowband spectrum-sensing techniques. In this technique, the input signal $x(t)$ is first taken from a channel, converted into a digital signal, processed using an $N$-point fast Fourier transform, and averaged over the $N$ sample points, as shown in Fig.~\ref{Fig2: Energy detection}. To decide whether the channel is vacant, we perform hypothesis testing. In this test, a predefined threshold $\lambda_{ED}$ is compared with the $N$-point average energy $E_{ED}$ of the $N$ samples. The output is a binary decision represented by (\ref{Eq: Ho}) and (\ref{Eq: H1}), indicating either hypothesis 0 ($H_0$) or hypothesis 1 ($H_1$). Here, $H_0$ denotes the absence of the primary user (PU) paying for channel services, while $H_1$ denotes the presence of the PU~\cite{arshid2022energy, luo2022energy}.
	
	\begin{equation}\label{Eq: Ho}
		E_{ED} < \lambda_{ED} \quad (H_0: \text{channel is vacant}),
	\end{equation}
	\begin{equation}\label{Eq: H1}
		E_{ED} > \lambda_{ED} \quad (H_1: \text{channel is busy}).
	\end{equation}
	
	In this study, we focus on two crucial parameters of the energy-detection channel-sensing technique: (i) the probability of successfully detecting a vacant channel, denoted by $P_d$, which represents the likelihood that the detector correctly detects the channel as vacant when it is actually vacant; and (ii) the probability of detecting the channel as busy when it is actually vacant, known as the probability of false alarm and denoted by $P_f$, which represents the likelihood that the detector incorrectly identifies a vacant channel as busy. The expressions for $P_d$ and $P_f$ are provided in (\ref{Eq: P_d}) and (\ref{Eq: P_f}), respectively~\cite{luo2022energy}.
	\begin{equation}\label{Eq: P_d}
		P_d = p(H_0\mid H_0) = Q\left(\frac{\lambda_{ED}-N(\sigma_n^2 + \sigma_s^2)}{(\sigma_n^2 + \sigma_s^2)\sqrt{2N}}\right),
	\end{equation}
	\begin{equation}\label{Eq: P_f}
		P_f = p(H_1\mid H_0) = Q\left(\frac{\lambda_{ED}-N\sigma_n^2}{\sigma_n^2\sqrt{2N}}\right),
	\end{equation}
	where $Q(\cdot)$ and $p(\cdot)$ represent the Q-function and probability function, respectively, and $\sigma_n$ and $\sigma_s$ represent the standard deviations of the noise and signal $x(t)$, respectively.
	
	In the proposed model, User1 senses the channel with probability $P_d$, which is treated as the BuC-detection probability, and then shares the BuC information with the eNB. After that, if User2 is also under the control of the same eNB, the sensed BuC can be used for communication, which establishes a controlled D2D link, as shown in Fig.~\ref{Fig1: System Model}.
	
	\section{Proposed Scheme}\label{Sec4: Proposed Work}
	This section describes the call-initiation flow diagram in CNs from the literature, which serves as the benchmark~\cite{nawaf2020reduce} for this work. The differences between the benchmark scheme and the proposed scheme are highlighted. Furthermore, we derive several mathematical expressions for the proposed scheme, which play a crucial role in understanding and evaluating its performance.
	
\begin{figure*}[htbp]
		\includegraphics[width=\linewidth]{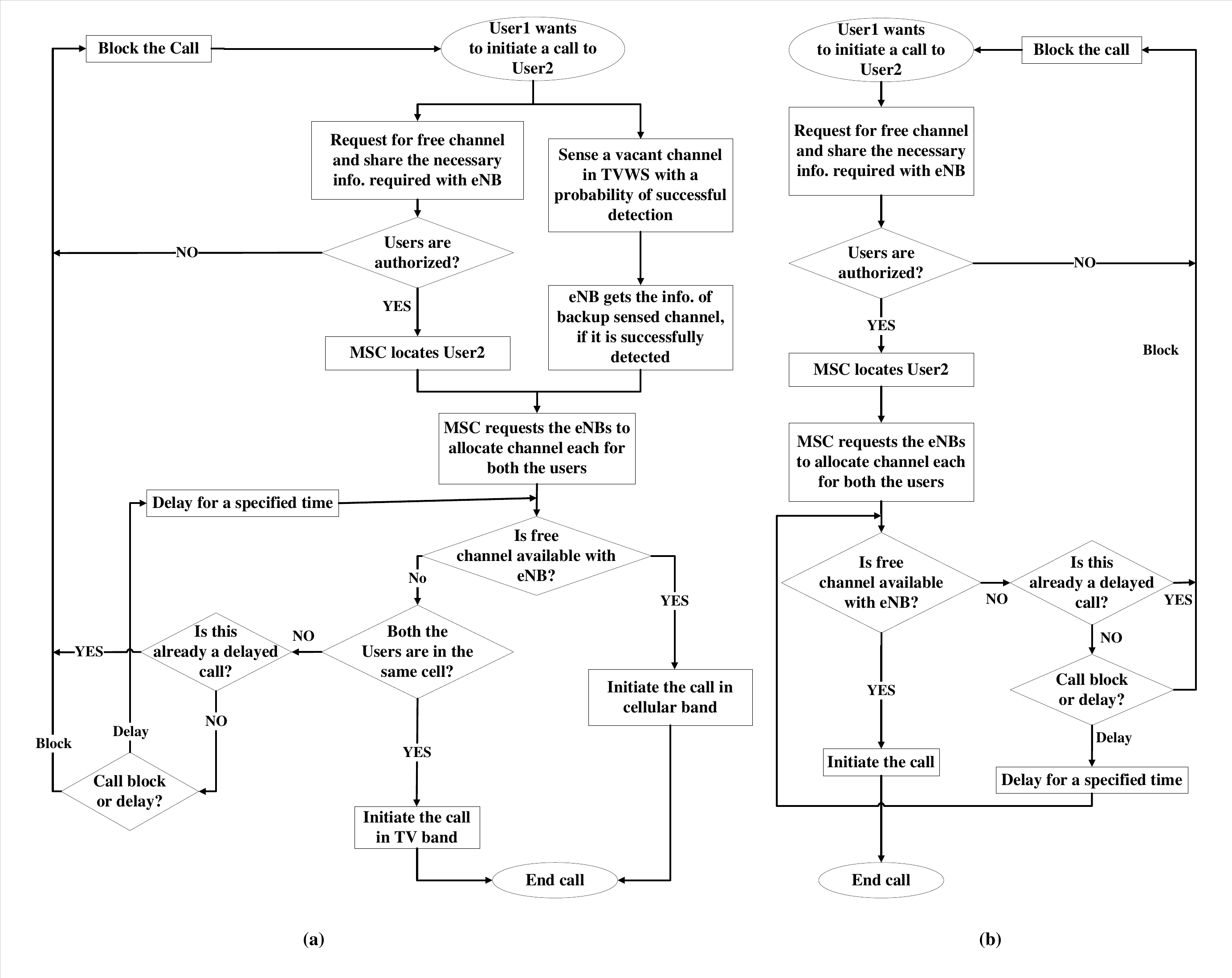}
		\caption{Flow diagrams illustrating call initiation in CNs from start to finish, comparing (a) the proposed model and (b) the conventional flow.}
		\label{Fig3: Flow Chart}
	\end{figure*}

	In the conventional call-initiation technique, when CU User1 wants to call another CU, User2, User1 sends all necessary information required for call initiation to its eNB. The eNB forwards the received information to the MSC to authenticate both CUs and find the location of User2. Then, the MSC requests the eNB/eNBs of both CUs to assign vacant channels for communication. If the eNB has vacant channels, it assigns them to the CUs. However, if there are no vacant channels, the call is instantly blocked or delayed for a certain amount of time. After a short delay, the eNB checks whether a channel is free. If no channel is free, the call is blocked. This phenomenon is illustrated in Fig.~\ref{Fig3: Flow Chart}(b).
	
	We propose a BuC sensed by the CU (User1) during call initiation. The channel-sensing process occurs in parallel with the processes that occur at the MSC/core CN, as shown in Fig.~\ref{Fig3: Flow Chart}(a). In our proposed solution, we introduce a BuC in TVWS. The BuC helps establish a controlled D2D link between the two CUs (i.e., User1 and User2). If both CUs are in the same macro-cell under the coverage of a single eNB and the eNB has no vacant channel, the users can be connected through the BuC. The proposed model significantly improves the QoS of CNs in populated urban areas.
	
	Erlang formulas are employed to calculate the probability of call blocking (Erlang-B) and the probability of call delay (Erlang-C) in this study. Some important assumptions considered while using Erlang formulas are as follows:
	\begin{itemize}
		\item Call arrivals follow a Poisson distribution;
		\item There is an infinite number of CUs;
		\item Any CU, including a blocked CU, can request a channel at any time;
		\item Longer calls are less likely to occur, as described by an exponential distribution; and
		\item Each eNB of each cell has a finite number of available channels.
	\end{itemize}
	These assumptions are crucial for accurately applying Erlang formulas to our analysis of call-blocking and call-delay probabilities in the CN scenario.
	
	\subsection{Probability of Call Blocking}\label{Sub Sec: Pb}
	Consider a macro-cell with $U$ CUs, where the call duration of each CU is $H$ and the average call-request rate is $\lambda$. Then, the traffic generated by each user is $A_u$, and the total traffic intensity generated by $U$ users, denoted by $A$, is given in (\ref{Eq: A_u}) and (\ref{Eq: A}) as
	\begin{equation}\label{Eq: A_u}
		A_u = \lambda H,
	\end{equation}
	\begin{equation}\label{Eq: A}
		A = UA_u.
	\end{equation}
	Furthermore, the probability of call blocking according to the Erlang-B formula is given by
	\begin{equation}\label{Eq: Erlang-B}
		P_b = \frac{\frac{A^C}{C!}}{\sum_{i=0}^{C} \frac{A^i}{i!}},
	\end{equation}
	where $C$ is the number of channels allocated to each eNB and $A$ is the total network traffic in each cell.
	
	In our proposed scheme, $P_b$ is reduced significantly because we introduce a BuC for CUs located in the same macro-cell. Thus, in our model, the probability that User1 and User2 are in the same macro-cell, $p(U_1U_2)$, and the probability that the BuC is successfully detected as vacant, $P_d$, are subtracted from $P_b$, as given below:
	\begin{equation}\label{Eq: New blocking Prob.}
		P_B = P_b - P(U_1U_2\cap C_d),
	\end{equation}
	where $P_B$ is the call-blocking probability for the proposed scheme, and $U_1U_2$ and $C_d$ are the events that User1 and User2 are in the same macro-cell and that a BuC is successfully detected in the TVWS band, respectively. Thus, $P(U_1U_2\cap C_d)$ represents the mutual occurrence of both events. Furthermore, the events $U_1U_2$ and $C_d$ are independent. Hence, $P(U_1U_2\cap C_d)$ can be defined as follows:
	\begin{equation}
		P(U_1U_2\cap C_d) = p(U_1U_2)p(C_d).
	\end{equation}
	Here, $p(U_1U_2)$ and $p(C_d)$ are the probabilities that both users are in the same macro-cell and that a BuC is successfully detected as vacant, respectively. Furthermore, we define $p(C_d) = p(H_0\mid H_0) = P_d$ in (\ref{Eq: P_d}). Therefore, the final expression for the call-blocking probability used in our proposed scheme is given by
	\begin{equation}\label{Eq: final blocking}
		P_B = P_b - p(U_1U_2)P_d,
	\end{equation}
	where $P_b$ and $P_d$ are defined in (\ref{Eq: Erlang-B}) and (\ref{Eq: P_d}), respectively.
	
	\subsection{Probability of Call Delay}\label{Sub Sec: Pdelay}
	The probability of delaying a call for more than 0 seconds, $P_{\mathrm{delay}}[D>0]$, is given by the Erlang-C formula as
	\begin{equation}\label{Eq: Erlang C}
		P_{\mathrm{delay}}[D>0] = \frac{\frac{A^C}{C!}}{A^C+C!\left(1-\frac{A}{C}\right)\sum_{i=0}^{C-1} \frac{A^i}{i!}}.
	\end{equation}
	We are interested in delaying a call for a particular time $t$ when a free channel is not instantly available with the eNB. The authors in~\cite{rappaport1996wireless} use a conditional-probability approach and give the expression below:
	\begin{equation}\label{Eq: Erlang C T}
		\begin{split}
			P_{\mathrm{delay}}[D>t]
			&= \left(\frac{\frac{A^C}{C!}}{A^C+C!\left(1-\frac{A}{C}\right)\sum_{i=0}^{C-1} \frac{A^i}{i!}}\right) \\
			&\quad \times \exp\left(\frac{-(C-A)t}{A}\right).
		\end{split}
	\end{equation}
	The probability of call delay used in our proposed scheme is given in (\ref{Eq: New delay Prob.}) based on the discussion in Subsec.~\ref{Sub Sec: Pb}:
	\begin{equation}\label{Eq: New delay Prob.}
		P_{\mathrm{DELAY}}[D>t] = P_{\mathrm{delay}}[D>t] - p(U_1U_2)P_d,
	\end{equation}
	where $P_{\mathrm{DELAY}}[D>t]$ is the probability of call delay used in our proposed scheme, and $P_{\mathrm{delay}}[D>t]$ can be obtained from (\ref{Eq: Erlang C T}).
	
	\section{Simulation Results and Discussion}\label{Sec5: Simulation Results}
	In this section, we evaluate and compare the performance of our proposed scheme for CNs with the benchmark scheme in~\cite{nawaf2020reduce}. We carried out the simulation experiments using MATLAB R2021b. The initial network simulation parameters are listed in Table~\ref{tab: simulation parameters}.
	
	\begin{table}[htbp]
		\centering
		\caption{Simulation parameters}
		\label{tab: simulation parameters}
		\resizebox{\columnwidth}{!}{%
			\begin{tabular}{|>{\centering\arraybackslash}m{2cm}|>{\centering\arraybackslash}m{1.2cm}|>{\centering\arraybackslash}m{2.5cm}|>{\centering\arraybackslash}m{1.25cm}|}
				\hline
				\textbf{Simulation} & \textbf{Parameter} & \textbf{Description} & \textbf{Value} \\ \hline
				\multirow{4}{2cm}{\centering Call blocking and call delay} & $\lambda$ & Call arrival rate & 0.65 calls/s \\ \cline{2-4}
				& $H$ & Call holding time & 180 s \\ \cline{2-4}
				& $T_d$ & Maximum call delay & 3 s \\ \cline{2-4}
				& $C$ & Channels per cell & 20 \\ \hline
				\multirow{5}{2cm}{\centering CR energy-detection sensing} & $U_{CR}$ & TV band users in the cell & 15 \\ \cline{2-4}
				& $n$ & Number of simulation iterations & $5000$ \\ \cline{2-4}
				& $B$ & Bandwidth & $6\times10^{6}$ Hz \\ \cline{2-4}
				& $T_s$ & Sensing time & $1\times10^{-3}$ s \\ \cline{2-4}
				& $N$ & Number of samples & $T_s\times B$ \\ \hline
			\end{tabular}%
		}
	\end{table}
	
	\subsection{Probability of Call Blocking}
	Call blocking occurs when a free channel is not instantly available for CU call initiation with its eNB. In this case, the CU request for a free channel in the CN is completely blocked by the eNB. Thus, the call-initiation request is terminated instantly, and the CU can request a free channel again at any time.
	
\begin{figure}[htbp]
		\includegraphics[width=\linewidth]{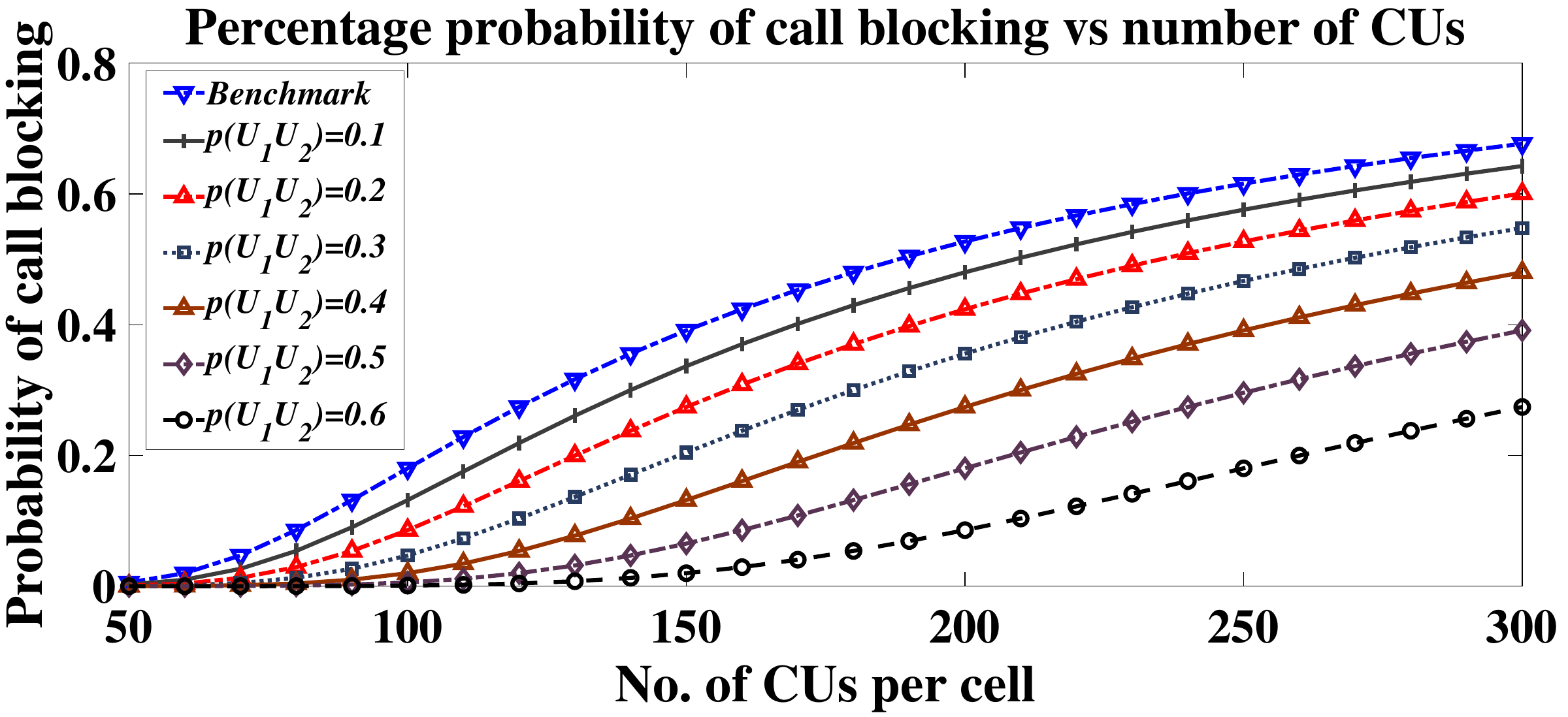}
		\caption{Comparison of call-blocking probability between the benchmark~\cite{nawaf2020reduce} and proposed scheme under varying network loads.}
		\label{Fig4: Blocking Prob.}
	\end{figure}

	The probability of call blocking versus the number of CUs in a cell is shown in Fig.~\ref{Fig4: Blocking Prob.}. Generally, the probability of call blocking increases as the number of CUs in the cell increases. In our proposed scheme, this probability is reduced because of the BuC in TVWS. The capacity of the cell is increased because more CUs can be served with the same resources. Furthermore, we assume that the probability of User2 being in the same macro-cell is 0.1, 0.2, 0.3, 0.4, 0.5, and 0.6, as shown in Fig.~\ref{Fig4: Blocking Prob.}. We observe that as the probability of both CUs being in the same cell increases, the probability of call blocking is significantly reduced, changing from an exponentially increasing curve to a nearly linear curve. This shows QoS improvement for congested, urban, and densely populated areas, such as international sports stadiums, marketplaces, and remote campuses, where the probability of both communicating CUs being in the same cell is relatively high.
	
	\subsection{Probability of Call Delay}
	Call delay occurs when the CU request for a free channel is delayed for a specified time $T_d$ due to the instant unavailability of a free channel with the eNB in CNs. After time $T_d$, the eNB checks whether a channel is free. It is more suitable to delay a CU request than to block it completely. When a CU request is blocked, the request must pass through the entire process again, whereas for a delayed request, the eNB checks for a free channel after time $T_d$, as shown in Fig.~\ref{Fig3: Flow Chart}.
	
\begin{figure}[htbp]
		\includegraphics[width=\linewidth]{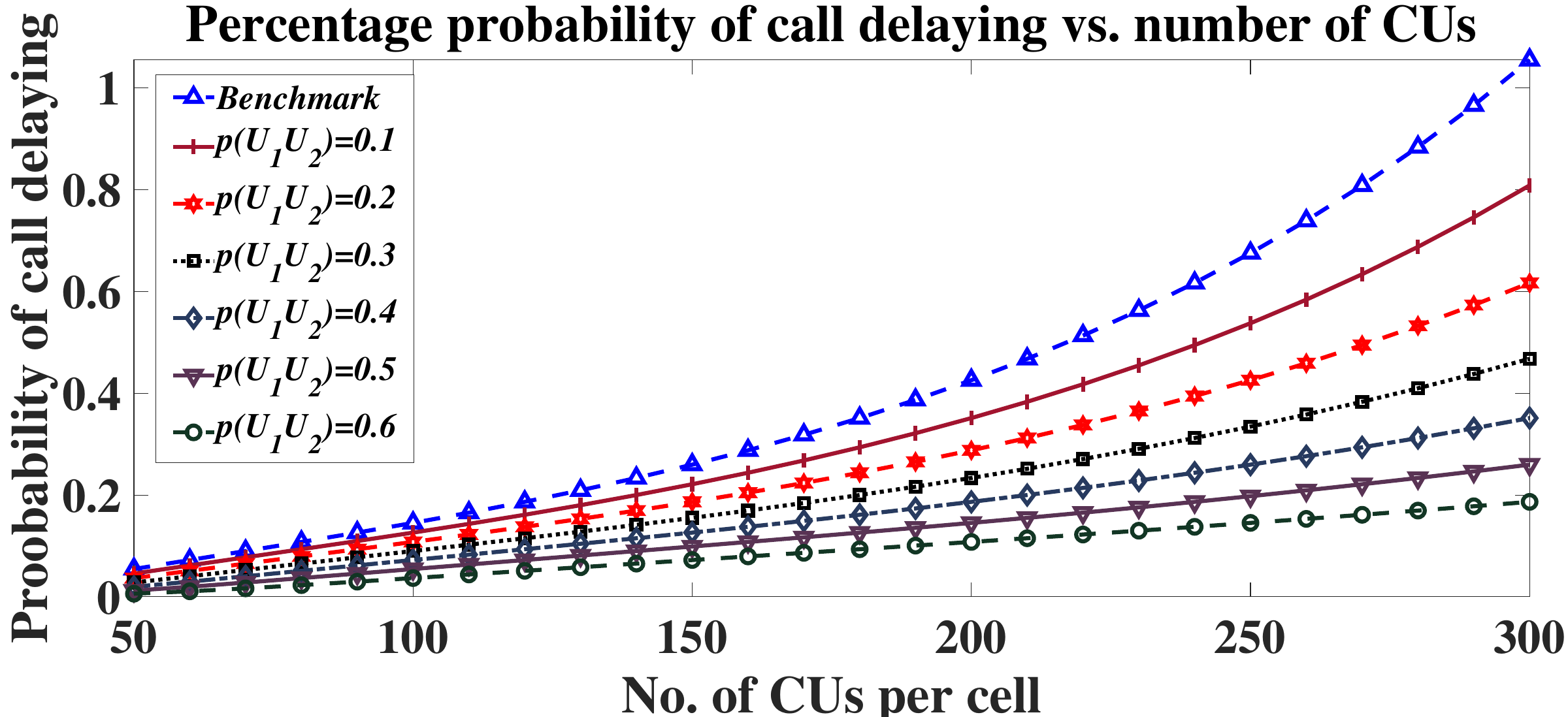}
		\caption{Comparison of the probability of call delay under different network-load conditions between the benchmark~\cite{nawaf2020reduce} and proposed scheme.}
		\label{Fig5: Delaying Prob.}
	\end{figure}

	The probability of call delay versus the number of CUs in a CN cell is shown in Fig.~\ref{Fig5: Delaying Prob.}. Generally, as the number of CUs in a cell increases, the probability of call delay increases. In our proposed scheme, because of the BuC in TVWS and the probability of both communicating CUs being in the same cell, the probability of call delay is significantly reduced. The simulation results show that when the probability of both communicating users being in the same cell increases, the graph changes from exponential to linear, as shown in Fig.~\ref{Fig5: Delaying Prob.}. The results can be further improved in populated environments, as discussed earlier.
	
	\section{Conclusions and Future Work}\label{Sec6: Conclusion}
	This research article addresses the problems of call blocking and call delay in CNs arising from limited spectrum resources with the eNB. To mitigate these issues, a novel approach is introduced in which a BuC is sensed using a CR energy-detection channel-sensing technique in TVWS. The benefits of the proposed approach include enhanced network capacity and reduced call-blocking and call-delay probabilities in congested, urban, and densely populated environments.
	
	The proposed work significantly improves CN performance, but certain challenges remain. For instance, UHF antennas for TV band communication may require larger sizes, and communication between multiple pairs of D2D users may encounter interference issues. Furthermore, other Internet of Things (IoT) devices may also utilize TVWS for communication in unlicensed bands, which can introduce interference. In future work, we will address these challenges by refining antenna usage, mitigating interference in multi-user D2D communication, and devising strategies to accommodate IoT devices operating in TVWS.
	
	\section*{Acknowledgment}
	The authors acknowledge the support of Najran University, Kingdom of Saudi Arabia, for this work.

\end{document}